\documentclass{svproc}
\usepackage{url}

\usepackage{xspace}
\newcommand{\pt}{\ensuremath{p_{\rm T}}\xspace}
\newcommand{\sqrts}{\ensuremath{\sqrt{s}}\xspace}
\newcommand{\rt}{\ensuremath{R_{\rm T}}\xspace}

\usepackage{cite}
\usepackage{caption}
\usepackage{graphicx}

\begin{document}
\mainmatter              % start of a contribution
\title{Particle production as a function \\ of underlying-event activity \\ measured with ALICE at the LHC}
\titlerunning{Particle production as a function of UE activity}  % abbreviated title (for running head)
\author{Valentina Zaccolo, on behalf of the ALICE Collaboration}
\authorrunning{V. Zaccolo for the ALICE Collaboration} % abbreviated author list (for running head)
%Craig Chambers, Kim B. Bruce, and Elisa Bertino}
%
\institute{University of Trieste and INFN, \\ Via Alfonso Valerio 2, 34127 Trieste,  Italy,\\
\email{valentina.zaccolo@ts.infn.it}}

\maketitle              % typeset the title of the contribution

\begin{abstract}

ALICE has performed several measurements aimed at understanding the collective-like effects observed in small collision systems.  
New approaches may be needed to clarify particle-production mechanisms in high-multiplicity pp collisions. 
Transverse momentum (\pt ) spectra as a function of charged-particle multiplicity show intriguing features. 
For example, data exhibit a stronger-than-linear increase of the self-normalised high-\pt\ particle yields versus multiplicity. 
In order to understand the role of auto-correlations on these effects, it has been proposed to use the underlying event as a multiplicity estimator to factorise the hardest and the softest components of the events. 
This approach can also be used to study collective effects in events with exceptionally large activity in the underlying-event region with respect to the event-averaged mean. 
In these proceedings, \pt\ spectra as a function of underlying-event activity in pp collisions measured with the ALICE detector are presented. 
Results are compared with PYTHIA 8.2 event generator.

\end{abstract}
\section{Introduction}
In hadronic interactions at high energies, as the ones achieved at the Large Hadron Collider (LHC), there are significant contributions from hard processes, which can be described by perturbative QCD precise calculations. 
Nevertheless, particle production at LHC is still dominated by soft-QCD processes.
Soft QCD is characterised by non-perturbative phenomenology and requires accurate modelling. 
The Underlying Event is constituted by multiple semi-hard parton interactions, initial and final state radiation and beam remnants, therefore, all the event components but the hardest scattering. 

The results presented in the following sections are motivated by two observations.
Firstly, the \pt\ spectra versus multiplicity results measured recently by ALICE~\cite{Acharya:2019mzb} highlighted a stronger than linear increase with multiplicity, which grows with \pt .
Using a forward multiplicity estimator, that is well separated in rapidity from the region of the measurement, the increase is reduced.
One could argue that the reason of the increase with multiplicity is due to the presence of jets which bias the estimator.
This last point can be understood better using a jet-free multiplicity estimator, which could then help characterising the correlation effects between low- and high-\pt\ particle production.
The second observation relates to multiplicity-dependent studies for heavy-flavours and high-\pt\ particle production~\cite{Abelev:2012rz, Adam:2015ota, Acharya:2019mzb}, where ALICE observes, again, a non-linear particle production increase with multiplicity.
This effect is often attributed to multiplicity saturation due to coherent hadronisation effects.
Anyway, recently, the non-linear particle production increase with multiplicity was explained with auto-correlations effects~\cite{Weber:2018ddv}. The authors argue that removing them would lead to a weaker-than-linear increase.

The measurements are performed with ALICE, one of the four main experiments at the LHC. 
It is constituted by 18 different detector systems and its peculiarities are a very high momentum resolution and excellent particle identification capabilities~\cite{Aamodt:2008zz}. 
The central barrel detectors are embedded in a solenoidal magnet of $B$ = 0.5 T nominal field.
The tracking and vertexing detectors used for results shown in the following are the Inner Tracking System (ITS) and the Time Projection Chamber (TPC).
The events used in the analysis were collected in 2016 proton--proton (pp) collisions at \sqrts = 13 TeV selected using a charged-particle signal coincidence in V0A and V0C arrays of scintillator counters.

\section{Analysis and results}
The analysis was performed considering two regions defined by the relative azimuthal angle with respect to the leading charged particle: $|\Delta\phi|=\phi-\phi_{\rm leading}$.
The toward and transverse regions are defined by $|\Delta\phi|<\pi/3$ and $\pi/3<|\Delta\phi|<2\pi/3$, respectively. 
The relative transverse activity classifier, \rt , is the self-normalised particle density in the transverse region~\cite{Martin:2016igp}.
An additional selection is done to characterise only the UE plateau region (the region of the jet pedestal): $5<p_{\rm T}^{\rm leading}<40$ GeV/$c$.
Several intervals of \rt\ were selected in order to distinguish between low and high UE activity. 
From PYTHIA 8 simulations, one expects that the \rt\ distributions exhibit a Kobe-Nielsen-Olesen scaling~\cite{Ortiz:2017jaz}. 
\begin{figure}[htbp]
   \begin{minipage}{0.5\textwidth}
   \centering
        \includegraphics[width=\textwidth]{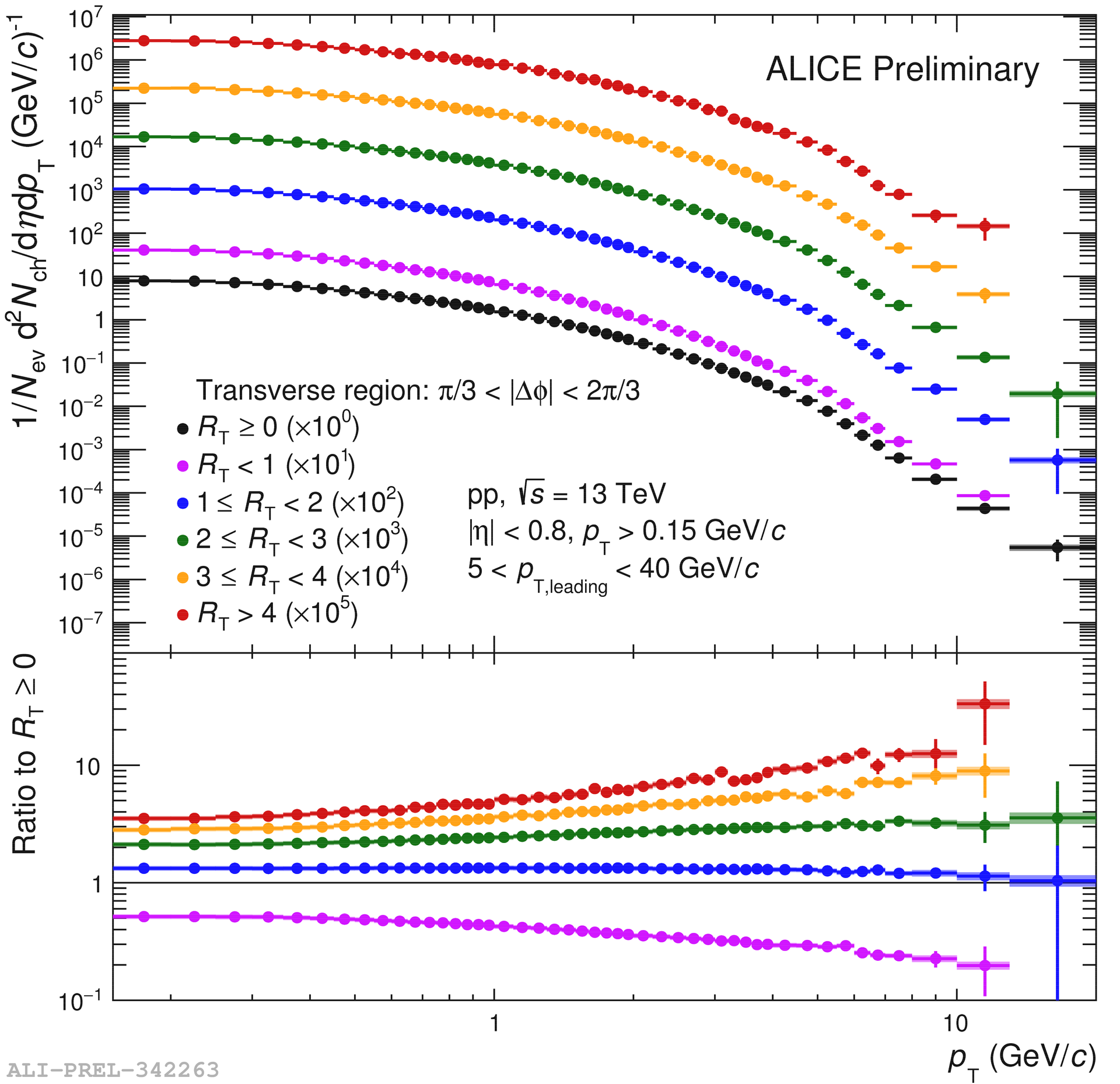}
   \end{minipage}
  \begin{minipage}{0.5\textwidth}
  \centering
        \includegraphics[width=\textwidth]{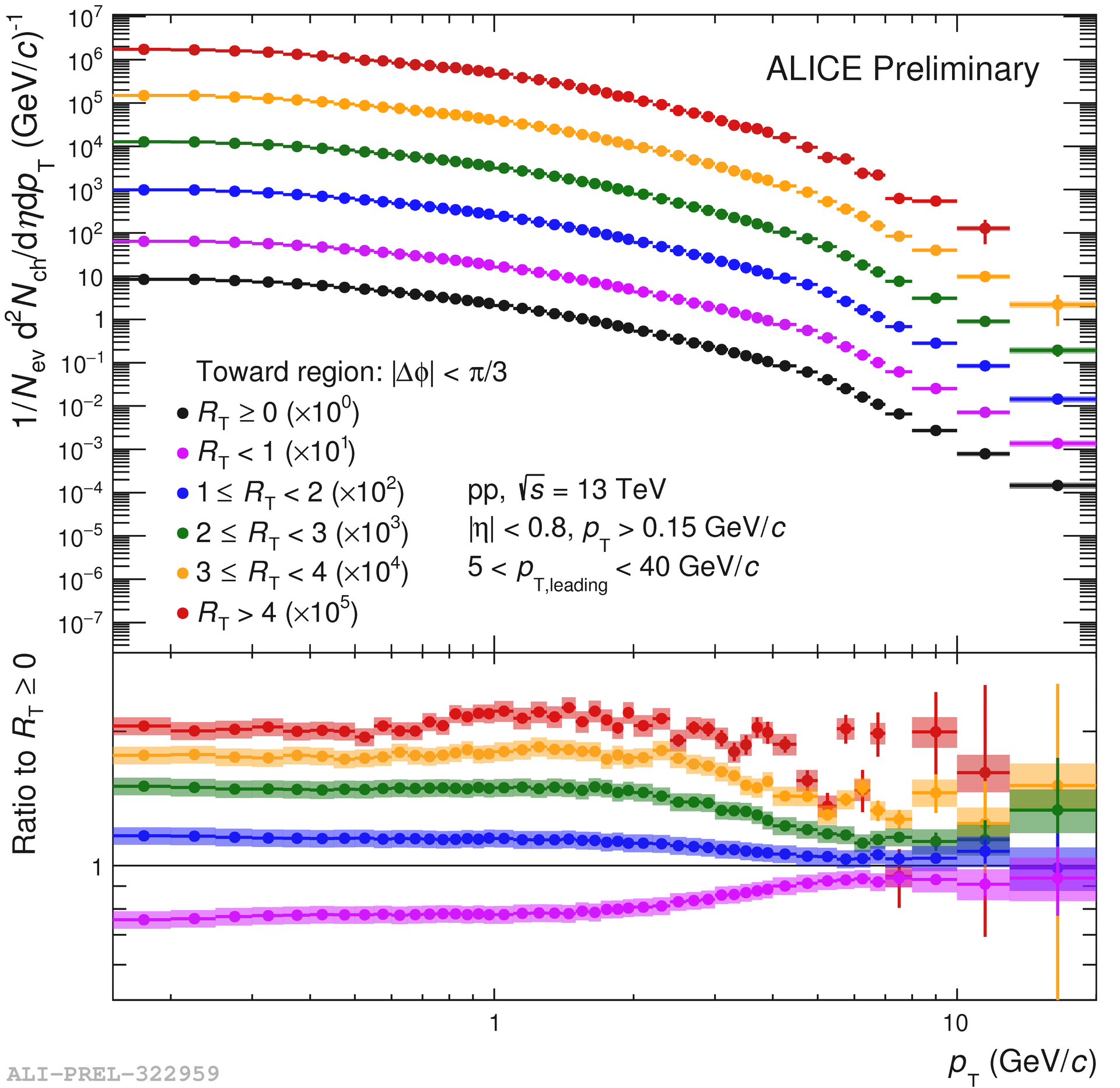}
	\end{minipage}
\caption{Charged-particle \pt\ spectra as a function of \rt\ in pp collisions at $\sqrt{s}=$ 13 TeV. Results for the transverse (left) and toward (right) regions are compared with their corresponding \pt spectrum for the \rt -integrated event class.  Vertical bars are statistical errors uncertainties, while boxes are systematic uncertainties.}\label{fig:1}	
\end{figure}
 
Results for the transverse region are shown in Fig.~\ref{fig:1} (left). 
The \pt\ spectra of charged particles reconstructed in the transverse region above 0.6 GeV/$c$ exhibit a strong \rt\ dependence. 
The evolution of the \pt -dependent spectra normalised to the inclusive \pt\ spectrum (\rt $\geq0$) are reminiscent of the observed behaviour when the inclusive multiplicity estimator was used~\cite{Acharya:2019mzb}. 
A different behaviour is observed for analogous measurements considering the toward region (Fig.~\ref{fig:1}, right), where the ratios to the inclusive \pt\ spectrum in the toward region converge to unity at high \pt . 
One can conclude that we have achieved an almost complete separation between the soft (UE) and hard (jet) parts of the event at high \pt .
Moreover, the auto-correlation effects are significantly reduced.

\begin{figure}[htbp]
    \centering
    \includegraphics[width=0.5\textwidth]{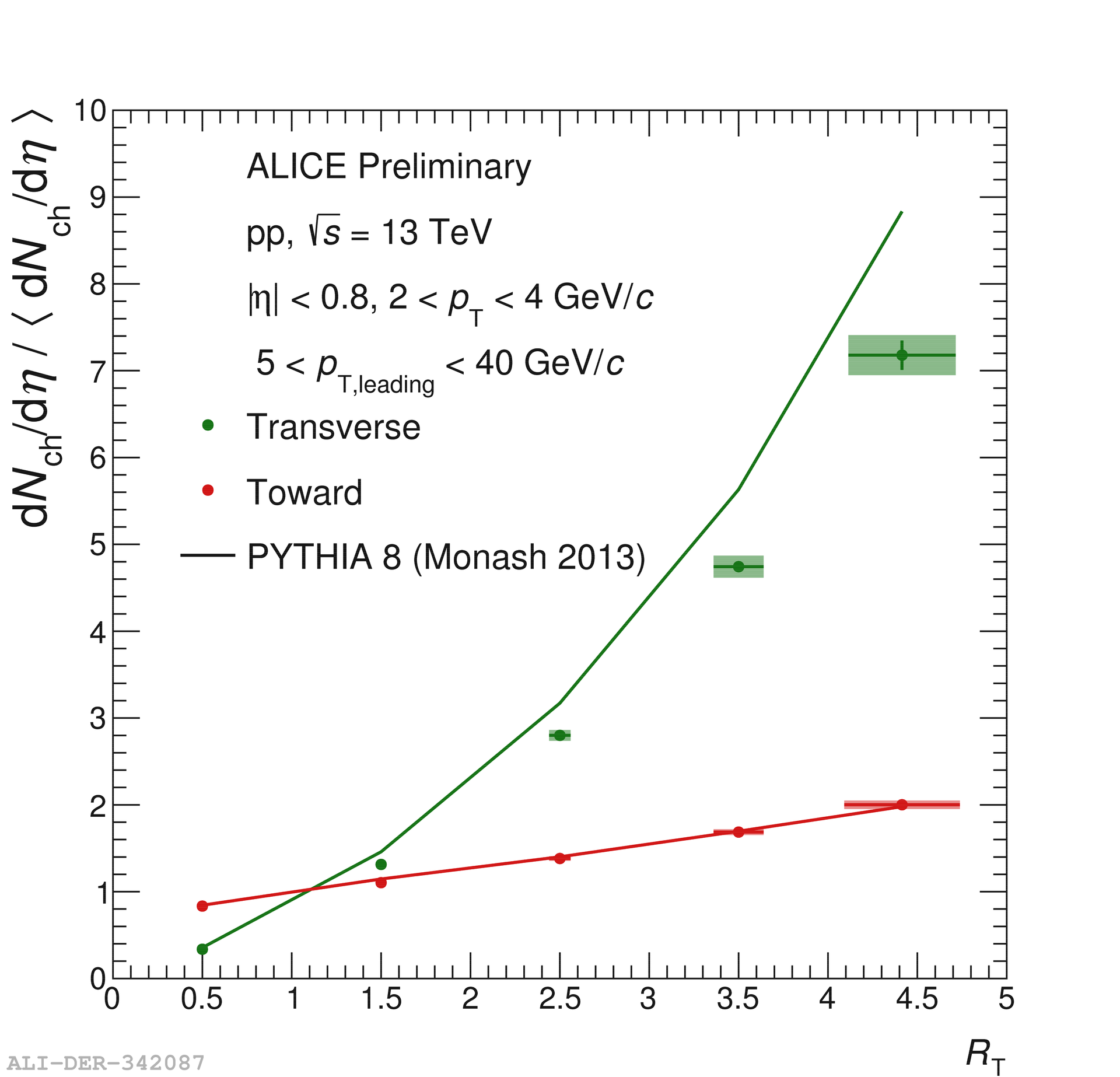}
\caption{Self-normalised charged-particle yields in the transverse (green) and toward (red) regions as a function of \rt\ for pp collisions at $\sqrt{s}=$ 13 TeV in the \pt\ range from 2 to 4 GeV/$c$. The data are compared with calculations of PYTHIA 8.2 with the Monash tune.}\label{fig:2}	
\end{figure}

Figure \ref{fig:2} shows the self-normalised particle yields, considering particles within $2<$ \pt\ $<4$ GeV/$c$, as a function of \rt . 
On one hand, the self-normalised yield in the transverse region, in green, shows the same behaviour of the one observed with the full midrapidity-based multiplicity estimator~\cite{Acharya:2019mzb}. 
On the other hand, the self-normalised yields in the toward region, in red, as a function of \rt\ do not converge to zero showing that at \rt $\sim$ 0 one can still have the presence of a jet.
This opens the possibility of studying a hard object with almost no UE activity.
Moreover, for the toward region the self-normalised yields exhibit a weaker dependence on \rt\ than that observed for the transverse region.
The results are compared to PYTHIA 8 (Monash tune) calculations~\cite{Sjostrand:2014zea} which reproduce well the observed trends.

\section{Summary and outlook}
The charged-particle \pt -spectra in the transverse and toward regions as a function of the relative transverse activity classifier \rt\ have been measured using the ALICE detector. 
Results for pp collisions at $\sqrt s$ = 13 TeV have been presented. 
The charged-particle \pt -spectra in the transverse region show a hardening at high \rt , which confirms the trend observed using the mid-pseudorapidity multiplicity estimator in the full azimuthal region. 
The charged-particle \pt -spectra in the toward region, instead, exhibit a weaker dependence with \rt , suggesting that the auto-correlation effects are at play when \pt\ spectra and multiplicity are both determined within the same pseudorapidity and $|\Delta\phi|$ regions. 
Finally, it was observed that it is still possible to identify a jet with zero activity in the UE, giving the opportunity to relate pp collisions to elementary colliding systems like $\rm e^{+}e^{-}$.
Therefore, \rt\ is an effective instrument to disentangle jet and UE components of the particle spectra and is a promising instrument for light and heavy-flavour particle-production studies.


\end{document}